\documentclass{scrartcl}
\usepackage{url}
\usepackage{graphicx}
\usepackage{amssymb}
\usepackage{color}
\usepackage{multicol}
\usepackage{xcolor}
  
\begin{document}
\title{Towards the Formalization of a Factory Demonstrator in BeSpaceD}

\author{
Keith Foster, Jan Olaf Blech, Guillaume Pr{\'e}vost
}
\date{RMIT University, Melbourne, Australia}

\maketitle
\begin{abstract}
This report gives an overview of our efforts towards a formalization for a food processing demonstrator
plant. Our BeSpaceD framework is used for the formalization. The
formalization comprises properties of components and relations between
components.  We present domain-specific constructs for the
formalization of industrial automation facilities and provide some
insights into the concrete food processing formalization. We are particularly interested in
spatio-temporal and other physical characteristics. Relationships are
formalized as graphs with annotated edges and components are
represented as nodes. 
\end{abstract}

\section{Introduction}
This report continues a series of documents \cite{b1,bf1,keith1} around our spatio-temporal modelling and
reasoning framework BeSpaceD\cite{bespacedURL}. In this report we  use the BeSpaceD language in a domain specific modelling task,
namely a factory demonstrator for food processing.
The models cover a variety of spatial and temporal aspects of the factory including physical location and sizing,
meta-data about each component, temporal correlations between actuators and sensors
as well as temporal safety margins between actuating.

BeSpaceD is our framework for spatio-temporal modelling and
reasoning\cite{bespaced1,bespaced0}.  BeSpaceD is implemented using Scala
and provides (i) spatio-temporal constructs for defining components
and their relationship especially with respect to time and space,
and (ii) other basic logic constructs to specify properties and models.
BeSpaceD is open source and the work described in this report is part
of the main development repository. The formalization presented here primarily serves academic
purposes and is a candidate for a reference formalization and example
for future efforts.

In addition to industrial automation \cite{coleng1,coleng2}, BeSpaceD has been applied to
mobile systems \cite{trains1,trains2} using the reactive blocks tool set \cite{reactive}, the verification of wireless network setups \cite{wireless} and in the area of smart energy systems \cite{smartspace,smartspace3d}. 
The work described in this paper makes use of qualitative spatial
representations \cite{cohn,chen} following ideas similar to principles
of the
region connection calculus \cite{rcc}. Initial ideas for the
formalization of the demonstrator that is subject to this report have
been discussed in \cite{harland}. The factory demonstrator is also
part of our VxLab activities \cite{vxlab}.

\subsection*{Overview}
We provide a short overview on BeSpaceD in Section~\ref{sec:bespaced}. Domain specific constructs for industrial automation are provided in Section~\ref{sec:ia} and Section~\ref{sec:demo} introduces our demonstrator. Section~\ref{sec:form} provides the formalization of the example system using BeSpaceD. A conclusion is featured in Section~\ref{sec:concl}.

\section{BeSpaceD}
\label{sec:bespaced}
The modelling constructs we use for the industrial automation domain are built on some generic data structures in the BeSpaceD core language. BeSpaceD provides some primitives for representing symbolic scalars and integers, component meta-data, events, time and space.
BeSpaceD builds most of its constructs on two foundational types:

{\small
\begin{verbatim}
    trait Invariant extends Ordered[Invariant]
    trait ATOM extends Invariant
\end{verbatim}
}

\noindent An Invariant is a formula that is supposed to hold for a system.
An ATOM is a special case of Invariant that is not composed of multiple invariants.
Time and space are of primary concern in BeSpaceD.
BeSpaceD represents {\tt time} with the following constructs:

{\small
\begin{verbatim}
    case class TimePoint [+T] (timepoint : T) extends ATOM
    case class TimeInterval[+T](timepoint1 : T, timepoint2 : T) extends ATOM
\end{verbatim}
}

\noindent BeSpaceD represents geometric {\tt space} with many occupational
constructs. Here, we primarily use:

{\small
\begin{verbatim}
    case class Occupy3DBox (x1 : Int, y1: Int, z1 : Int,
                            x2 : Int, y2 : Int, z2 : Int)
         extends ATOM
\end{verbatim}
}
BeSpaceD represents {\tt events} with a single construct called Event:

{\small
\begin{verbatim}
    case class Event [+E] (override val event  : E) extends ATOM
\end{verbatim}
}

\noindent For our modelling effort in industrial automation we extend
the the BeSpaceD language in a number of areas. The rest of this
section presents these extensions.

\subsection {Controller Events}

BeSpaceD represents events with a single construct called {\tt Event}, and a generalization called {\tt EventLike}:

{\small
\begin{verbatim}
    trait EventLike  [+E] extends ATOM { val event: E }
    case class Event [+E] (override val event  : E) extends EventLike[E]
\end{verbatim}
}

\noindent In the industrial automation domain an event is something that changes state at an instant in time.
There is a construct in BeSpaceD to reflect these semantics:

{\small
\begin{verbatim}
    class INSTATE[+O, +T, +S] (val owner : O, val timepoint : TimePoint[T],
                               val state :S)
          extends Event[(O,TimePoint[T],S)]( (owner, timepoint, state) )
\end{verbatim}
}

\noindent The value passed to the superclass Event's constructor is a tuple of cardinality three which maintains compatibility with the more general {\tt Event}.

\subsection {Structural Extensions}
We define meta-data for each of the components that make up a machine. This data lends itself to a key-value style encoding. In order to support this style we added the following BeSpaceD structural constructs to tag data with semantics in the context of this key-value encoding style:

{\small
\begin{verbatim}
    case class Component[+I] (id : I) extends ATOM
    case class ComponentValue[+V] (value : V) extends ATOM
\end{verbatim}
}

\noindent {\tt Component} defines the keys and {\tt ComponentValue} defines the values.
By convention, the keys identify a single component.
The type could be a string or a reference to any domain specific component.
In our case the component ID's are stings used to define a unique name for each.

Meta data is defined in BeSpaceD as a conjunction (BIGAND) of key value pairs,
where each pair is an implication (IMPLIES).
This is very similar to a "Map" in mathematics and many programming languages.
BeSpaceD provides the {\tt BeMap} construct (a specialization of BIGAND)
which extends the relatively simple BIGAND with map like behaviour:

{\small
\begin{verbatim}
    class BeMap[+P <: Invariant, +C <: Invariant](terms: List[IMPLIES[P,C]])
        extends BIGAND[IMPLIES[P,C]](terms)
        {
          lazy val premises    = ( this.terms map { _.premise } ).toSet[Any]
          lazy val conclusions = ( this.terms map { _.conclusion } ).toSet[Any]
          lazy val elements    = premises union conclusions
    
          def apply(key: Invariant): Option[C]
        }
\end{verbatim}
}

\noindent BeMaps are used to construct the descriptions using {\tt Component} as the key and {\tt Component\-Value} as the value:

{\small
\begin{verbatim}
    case class ComponentDescription[+C <: Component[Any],
                                    +V <: ComponentValue[Any]]
               (override val terms: List[IMPLIES[C,V]])
         extends BeMap[C,V](terms)
\end{verbatim}
}

\noindent An example construction of meta data is shown below:

{\small
\begin{verbatim}
    val elon = BeMap(
        Component("Name")    ==> ComponentValue("Elon Musk"),
        Component("Address") ==> ComponentValue("Mars")
        )
\end{verbatim}
}

\noindent An example query of a BeMap can be done as follows:

{\small
\begin{verbatim}
    elon("Address")
\end{verbatim}
}

\subsection {Topological Extensions}
To support the use of graphs for defining topologies we added a graph construct to BeSpaceD. 
In our model the edges are annotated so the existing BeSpaceD Edge primitive is extended to support annotations of any type:

{\small
\begin{verbatim}
    class EdgeAnnotated[+N, +A] (val source : N, val target : N,
                                 val annotation: Option[A])
          extends ATOM
    case class Edge[+N] (override val source : N, override val target : N)
         extends EdgeAnnotated[N,{\tt Nothing}](source, target, None)
\end{verbatim}
}

\noindent It is interesting to note, in type inheritance terms, {\tt Edge} is a specialisation of {\tt Edge\-Annotated}.
On the surface, this may feel counter-intuitive, however, it makes sense when you consider that {\tt Edge} proper is a special kind of edge annotated with nothing.
Scala provides a special type called {\tt Nothing} that is a specialisation of all other types and we make use of this feature.

The Edge and EdgeAnnotated constructs define edges with two nodes.
We define graphs as a BIGAND of annotated edge objects called {\tt BeGraphAnnotated}.
The non-annotated version is called a "BeGraph" for behavioural graph.

{\small
\begin{verbatim}
    class BeGraphAnnotated[+N, +A](terms: List[EdgeAnnotated[N, A]])
          extends BIGAND[EdgeAnnotated[N, A]](terms)	
    class BeGraph[+N](terms: List[Edge[N]])
          extends BeGraphAnnotated[N, Nothing](terms)
\end{verbatim}
}

\noindent An example construction of a simple graph would be:

{\small
\begin{verbatim}
    BeGraph(
           Component("Smoke Detector") --> Component("Alarm"),
           Component("Motion Sensor") --> Component("Alarm")
           )
\end{verbatim}
}

\noindent The {\tt-->} symbol is a binary infix operator that constructs an {\tt Edge} object.

\subsection {Temporal Extensions}
We model a number of different topologies for the industrial automation domain
most of which require temporal semantics.
We extended BeSpaceD with two temporal constructs to represent timing relationships:
correlations and constraints.
A temporal correlation declares that two events coincide with a specified time delay relative to the first event.
The first event is called the "cause" and the second, "effect".

{\small
\begin{verbatim}
    case class TemporalCorrelation[+E, +T <: Number]
         (cause: EventLike[E], duration: TimeDuration[E], effect: EventLike[E])
         extends Invariant
\end{verbatim}
}

\noindent A temporal constraint declares that two events (or one event and the lack of a second event in the inverse case) occur
within a specified time duration relative to the first event.

{\small
\begin{verbatim}
case class TemporalConstraint [+E, +T <: Number]
           (cause: EventLike[E], range: TimeDurationRange[E],
            effect: EventLike[E], inverse: Boolean = false)
     extends Invariant
\end{verbatim}
}

\section{Domain Specific Modelling for Industrial Automation}
\label{sec:ia}

We identified many data structures that are reusable within the industrial automation  domain
which are kept separated from applications.
For the modelling exercise we considered programmable logic
controllers (PLC) outside the scope.
The industrial automation domain meta-model is defined in the {\tt BeSpaceDIA} project and defines types which, in effect, present design patterns for developers of more specific domains such as food processing.
In this section we categorize and  summarize all the domain specific types in the {\tt PhysicalModel} module of  {\tt BeSpaceDIA}.

\subsection{Object Identity}
We provide a trait for each kind of physical object and then break this down into a generalised type hierarchy representing components of automation equipment.

{\small
\begin{verbatim}
    trait PhysicalObject[+I]       extends Component[I]
\end{verbatim}
}

A device represents any built in component of the machine itself.
This is further broken down into parts, actuators and sensors.
A {\tt Part} represents a device that forms part of a useful process.
An actuator is a device that connects PLCs with parts to make them do something.
A sensor is a device that communicates some portion of the state of the machine back to the PLC.
  
{\small
\begin{verbatim}
    trait Device[+I]               extends PhysicalObject[I]
    trait Part[+I]                 extends Device[I]
    trait Actuator[+I]             extends Device[I]
    trait Sensor[+I]               extends Device[I]
\end{verbatim}
}

{\tt Material} represents physical things that pass through the machine but are not considered part of the machine itself.
A discrete material represents atomic objects that should not be divided; for example bottle caps.
There is no point it measuring fractions of a cap, so these can be measured using natural numbers.
An analog material represents a divisible material where we can measure it using a real number; for example, water.

{\small
\begin{verbatim}
    trait Material[+I]             extends PhysicalObject[I]
    trait DiscreteMaterial[+I]     extends Material[I]
    trait AnalogMaterial[+I]       extends Material[I]
          { val unit: String; val quantity: Double }
\end{verbatim}
}

All physical objects have an identity.
With regards to discrete materials there is the ability to uniquely identify each one,
for example each bottle cap can have a unique identity.
This would allow one to trace discrete materials through the system.

\subsection{Spatial Variation}
The following type is defined for creating an application specific
spatial variation:

{\small
\begin{verbatim}
    trait SpatialVariation extends Invariant
\end{verbatim}
}

\noindent This allows the applications to model a set of discrete values representing positions in space (like an enumerated type).
This is done in Scala using singleton objects which extend the {\tt
  SpatialVariation} trait, for example:

{\small
\begin{verbatim}
    abstract class StackEjectorPosition extends ATOM with SpatialVariation
    object StackEjectorRetractedPosition extends StackEjectorPosition
    object StackEjectorExtendedPosition extends StackEjectorPosition
\end{verbatim}
}

\subsection{State Types}
\label{ssec:ia-var}
The following types are defined for creating application specific
states:

{\small
\begin{verbatim}
    trait PhysicalState[+S]           extends ComponentState[S]
    trait DeviceState[+DS]            extends PhysicalState[DS]
    trait PartState[+PS]              extends DeviceState[PS]
    trait ActuatorState[+AS]          extends DeviceState[AS]
    trait SensorState[+SS]            extends DeviceState[SS]
\end{verbatim}
}

\noindent This allows the applications to model a set of discrete values representing all the useful abstract states for each category of device, for example:

{\small
\begin{verbatim}
    abstract class LightSensorState(sig: Signal) extends SensorState(sig)
    class ObstructedState(sig: Signal) extends LightSensorState(sig)
    val ObstructedHigh   = ObstructedState(High)
    val ObstructedLow    = ObstructedState(Low)
    val Obstructed       = ObstructedState(DC)
\end{verbatim}
}

\noindent {\tt Obstructed} is an abstract state and {\tt ObstructedHigh} additionally records the actual domain specific electrical signal that indicates this state.

\subsection{Events}
\label{ssec:ia-event}

The following types are defined for creating application specific events:

{\small
\begin{verbatim}
    abstract class PhysicalEvent[+I, S](component: PhysicalObject[I],
                                        timepoint: TimePoint[Long],
                                        state: PhysicalState[S])
             extends INSTATE[PhysicalObject[I], Long, PhysicalState[S]]
                     (component, timepoint, state)
             with BeSpaceDGestalt.core.Event[S]
             
    abstract class DeviceEvent[+I, S](device: Device[I], 
                                      timepoint: TimePoint[Long],
                                      state: DeviceState[S])
             extends PhysicalEvent[I, S](device, timepoint, state) 

    abstract class ActuatorEvent[+I, S](actuator: Actuator[I],
                                        timepoint: TimePoint[Long],
                                        state: ActuatorState[S])
             extends DeviceEvent[I, S](actuator, timepoint, state) 
             
    abstract class SensorEvent[+I, S](sensor: Sensor[I],
                                        timepoint: TimePoint[Long],
                                        state: SensorState[S])
             extends DeviceEvent[I, S](sensor, timepoint, state) 
\end{verbatim}
}

\noindent The following allows the application to model live events describing changes in the state of the factory, for example:

{\small
\begin{verbatim}
    SensorEvent(StackEmpty, TimePoint(new Date()), Obstructed)
\end{verbatim}
}

\noindent This means the relevant light sensor has detected the stack is empty.

%\subsection{Relationships}
% Not Used

\subsection{Topologies}
\label{ssec:ia-top}
The following types are defined for creating application specific
topologies:

{\small
\begin{verbatim}
    abstract trait Relationship
    abstract trait SpatialRelationship extends Relationship
    abstract trait TemporalRelationship extends Relationship
    abstract trait SpatioTemporalRelationship
             extends SpatialRelationship with TemporalRelationship

    type Topology[+COM <: Component[Any], +REL <: Relationship] =
         BeGraphAnnotated[COM, REL]
    type SpatialTopology[+COM <: Component[Any],
                         +SREL <: SpatialRelationship] =
         BeGraphAnnotated[COM, SREL]
    type TemporalTopology[+COM <: Component[Any],
                          +TREL <: TemporalRelationship] =
         BeGraphAnnotated[COM, TREL]
    type SpatioTemporalTopology[+COM <: Component[Any],
                                +STREL <: SpatioTemporalRelationship] =
         BeGraphAnnotated[COM, STREL]
\end{verbatim}
}

\noindent The following allows the applications to model functional, spatial and/or temporal relationships between devices, for example:

{\small
\begin{verbatim}
        case class Seconds(s: Int)
             extends TimeDuration(s) with SpatialRelationship

        BeGraphAnnotated (
            EdgeAnnotated(Component("Smoke Detected"),
                          Component("Alarm Activated"),
                          Some(Seconds(1))),
            EdgeAnnotated(Component("Smoke Clear"),
                          Component("Alarm Deactivated"),
                          Some(Seconds(3)))
            )
\end{verbatim}
}

\section{The Example Factory Demonstrator}
\label{sec:demo}\

The food processing factory used in our demonstrator is scaled down version of a fully functioning factory system capable of
    filling bottles with liquid or corn kernels,
    capping the bottles, moving bottles on conveyer belts and
    supplying both bottles and caps from storage.

% Picture of the Station 1 here.
\begin{figure*}
    \centering
    \includegraphics[width=0.65\textwidth] {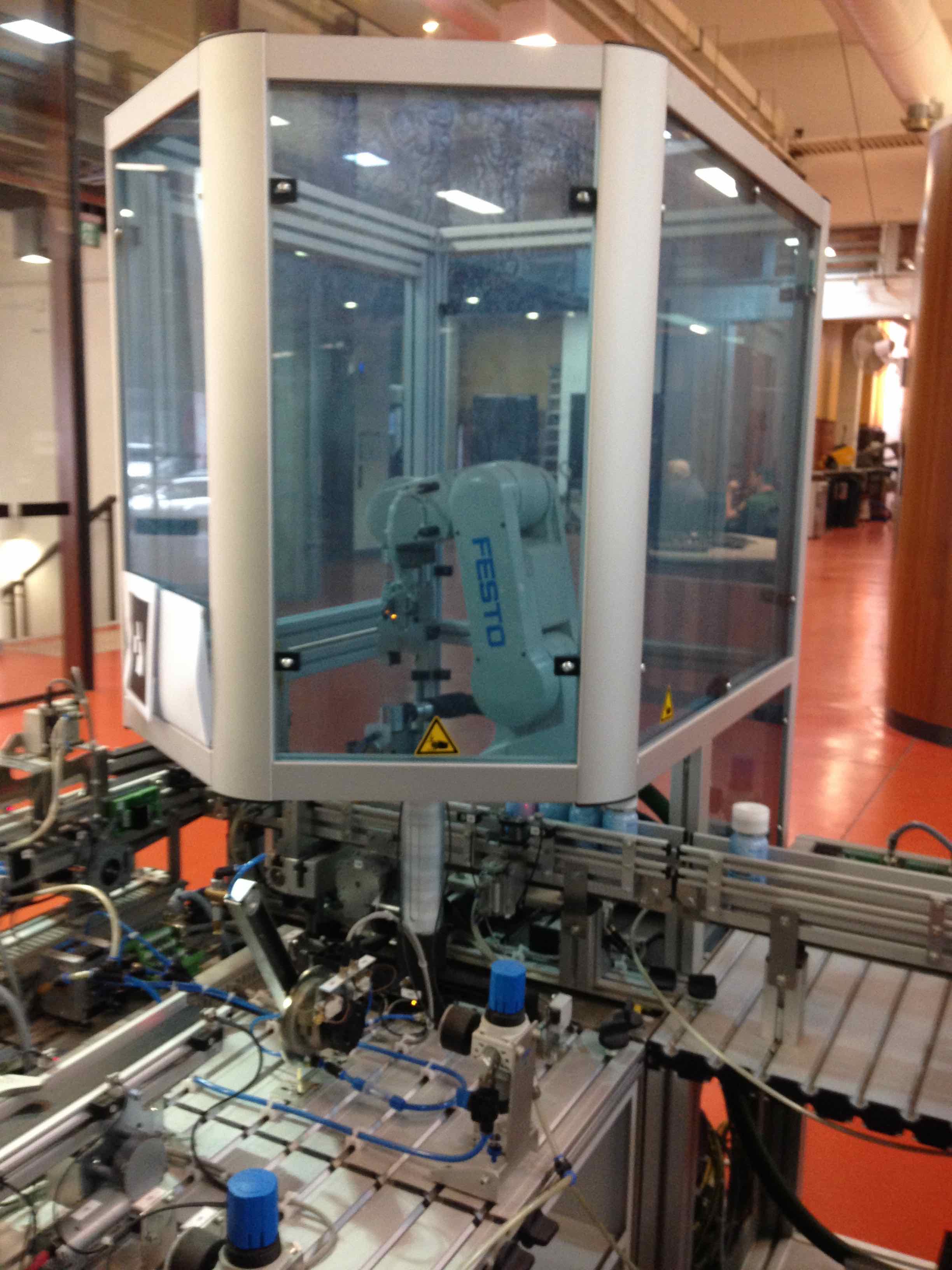}
    \caption{Station one}
    \label{fig:festo-station-1}
\end{figure*}

\subsection{Scope}
For this modelling application we limit the scope to a single station (as seen in Figure~\ref{fig:festo-station-1}) which is controlled by a single PLC.
The originally provided PLC is replaced with a customized Raspberry Pi based PLC solution which allows us to
run BeSpaceD and communicate to both the station and the cloud over a wifi network.
We numbered this Station one and named it ``CapDispenser''.
Other stations that are out of scope but we have partially modelled include: CapConveyer, Capping, RotaryTablePart1, RotaryTablePart2, CappedBottlesConveyerBelt, PackagingStation, BottleConveyer and BottleConveyerSorting.

\begin{figure*}
    \centering
    \includegraphics[width=0.5\textwidth] {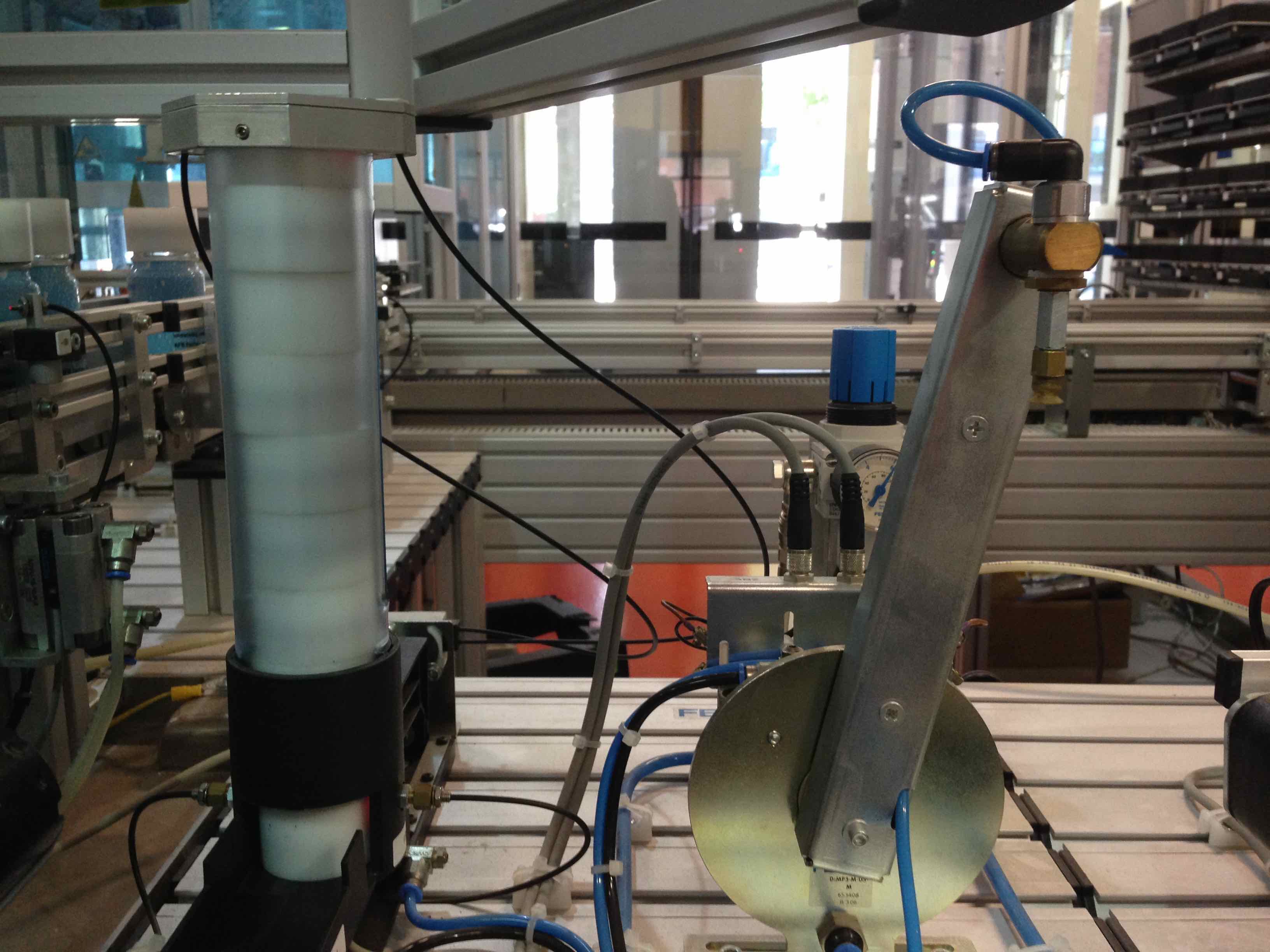}
    \caption{Two subsystems of station one}
    \label{fig:festo-stack-loader}
\end{figure*}

The CapDispenser Station involves two subsystems (as seen in Figure~\ref{fig:festo-stack-loader}).
{\small
\begin{itemize}
\item A tube to hold caps with a cap ejector underneath that pushes caps out of the tube (see Figure~\ref{fig:festo-stack}).
% Close-up picture of the tube and ejector.
\begin{figure*}
    \centering
    \includegraphics[width=0.5\textwidth] {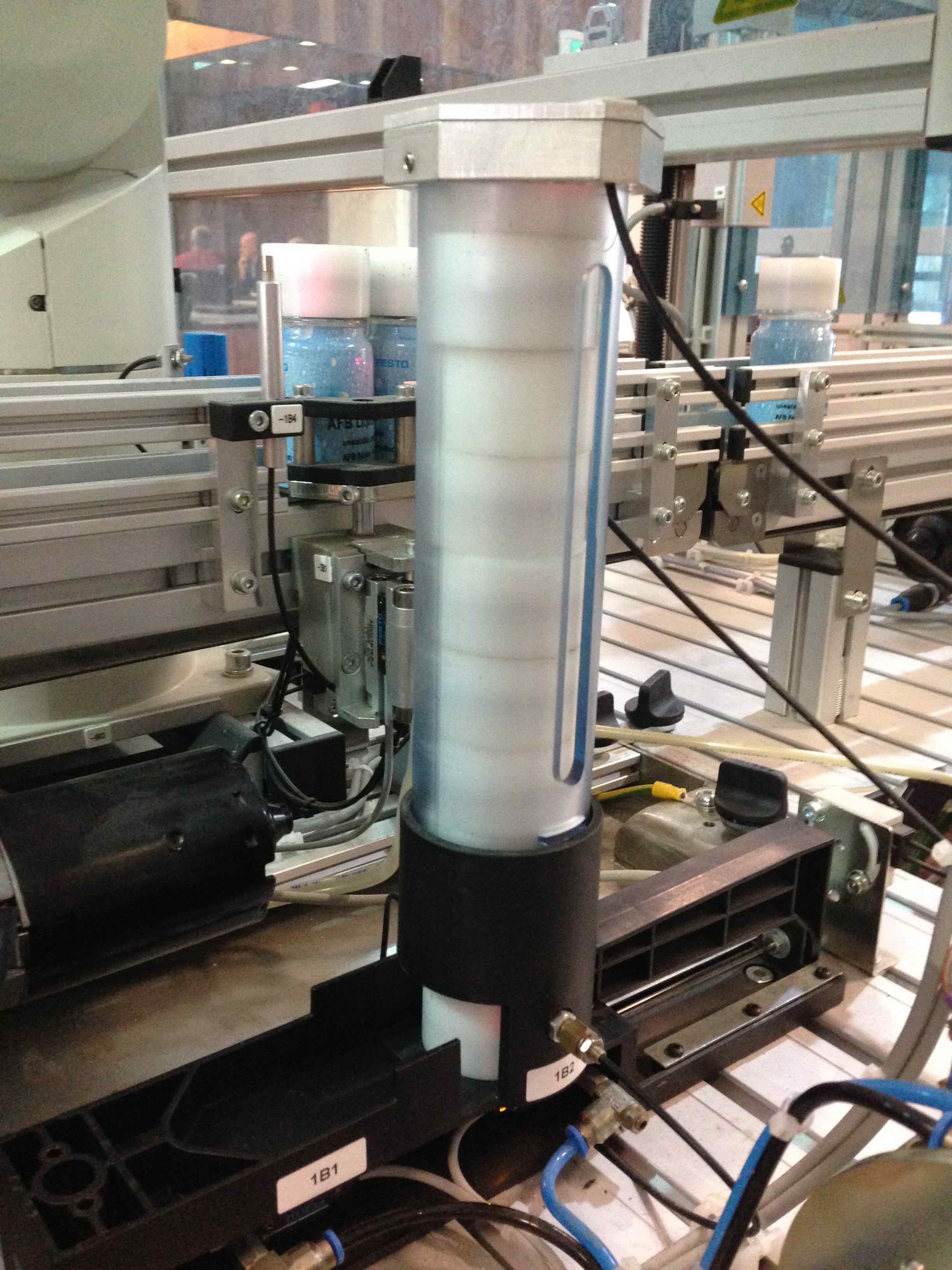}
    \caption{Cap stack and ejector}
    \label{fig:festo-stack}
\end{figure*}

\item A lever arm (that swings in an arc) with a vacuum gripper built into the tip of the arm used to pick up and drop off caps (see Figure~\ref{fig:festo-loader}).
% Close-up picture of the loader arm.
\begin{figure*}
    \centering
    \includegraphics[width=0.5\textwidth] {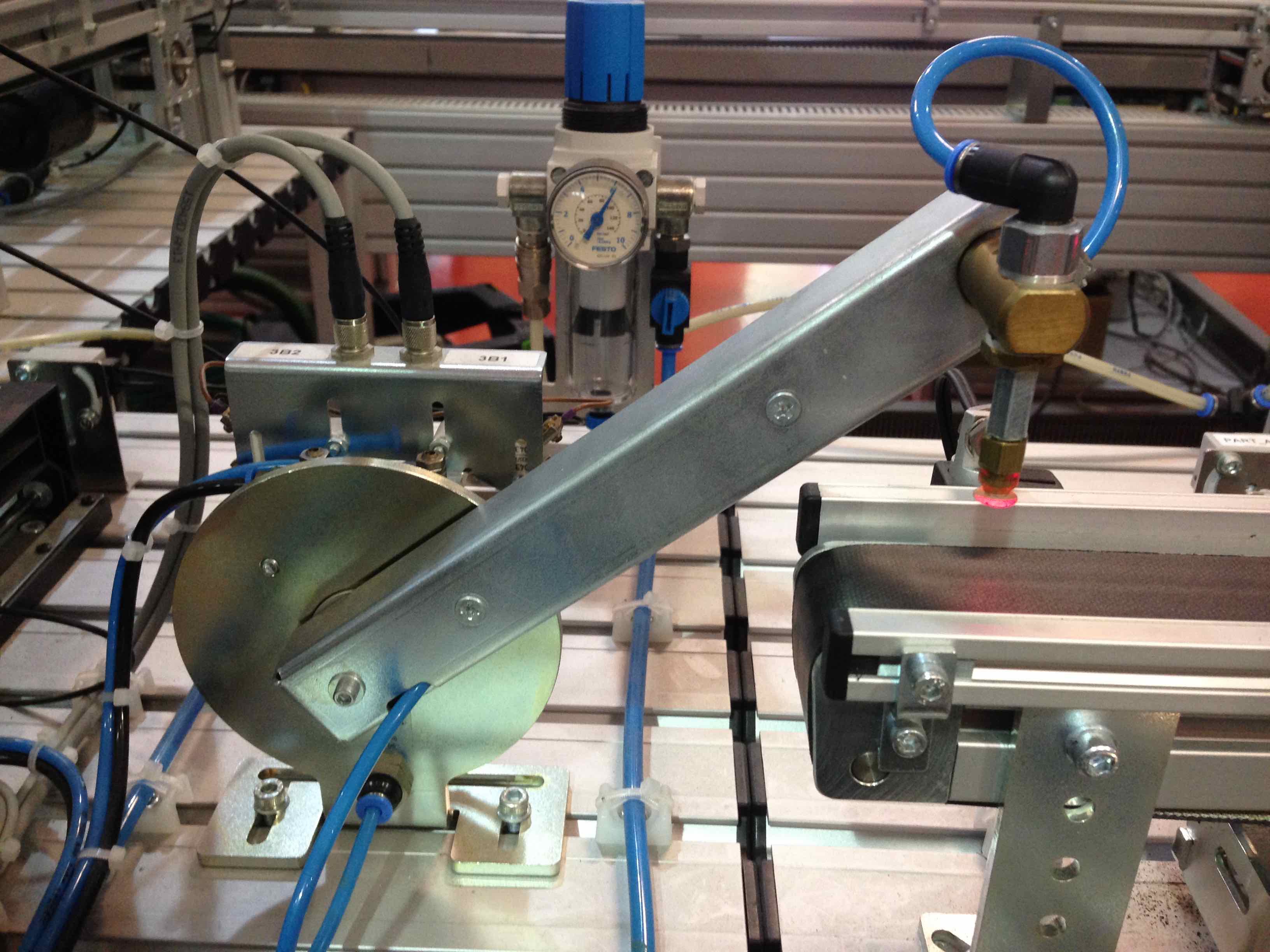}
    \caption{Lever arm with vacuum gripper}
    \label{fig:festo-loader}
\end{figure*}

\end{itemize}
}

\subsection{Description of the Cap Dispenser Station}
In this section we describe the two subsystems of the Cap Dispenser station in more detail including the parts, actuators and sensors.

\subsubsection{The Cap Stack and Ejector}

There is only one actuator integrated into the cap ejector part: 
{\small
\begin{itemize}
\item Stack Ejector Extend: Activation causes the ejector to be extended which pushes out the bottom cap in the stack if there is one.
\item Stack Ejector Extend: Inactivation causes the ejector to be retracted which returns the ejector to the original position and causes the rest of the caps to move down one cap height in the stack due to gravity.
\end{itemize}
}

There are three sensors integrated into the cap tube and ejector parts: 
{\small
\begin{itemize}
\item Stack Empty: A light sensor positioned at the base of the tube that when unobstructed indicates the tube is empty.
\item Stack Ejector Extended: A light sensor positioned at the extreme extended position of the ejector. When obstructed this indicates the stack ejector has been fully extended.
\item Stack Ejector Retracted: A light sensor positioned at the extreme retracted position of the ejector. When obstructed this indicates the stack ejector has been fully retracted.
\end{itemize}
}

\subsubsection{Pick-and-Place Unit}

There are four actuators integrated into the cap loader and vacuum gripper parts: 
{\small
\begin{itemize}
\item Loader Pickup: Activation causes the arm to rotate to the left towards the pickup area.
\item Loader Dropoff: Activation causes the arm to rotate to the right towards the drop-off area.
\item Vacuum Grip: Activation causes the vacuum motor to pump air out of the suction cup mounted in the tip of the lever arm.
\item Eject Air Pulse: Activation causes the pulse motor to pump air into the suction cup causing the cup to loose suction.
\end{itemize}
}

There are three sensors integrated into the cap loader and vacuum gripper parts: 
{\small
\begin{itemize}
\item Loader Picked Up: A contact sensor that indicates the arm has fully rotated into the pickup area.
\item Loader Dropped Off: A contact sensor that indicates the arm has fully rotated into the drop-off area.
\item Workpiece Gripped: A sensor that indicates the vacuum has reached a level where suction force is greater than the gravitational force of a bottle cap (i.e. it is gripped and should not fall off).
\end{itemize}
}

\section{Demonstrator Formalization}
\label{sec:form}

We divide our model of the food processing factory domain into four distinct areas:
{\small
\begin{itemize}
\item Components - For modelling the identity of each physical component in the factory. For this domain we call the components ``Devices''.
\item Meta data - For modelling static data about the components. For this domain we call this data ``Device Descriptions''. Spatial data is included in the descriptions. 
\item Topologies - For modelling information in a graph structure. For this domain we experimented with a few topologies representing different information.
\item Events - For modelling Dynamic data (i.e. data that is defined in time).
\end{itemize}
}

Each area is described in more detail below including definitions of the data structures used and some example code snippets.

\subsection{Devices (Components)}
The model of devices provides a uniform way to identify every component in the factory. As per the industrial automation domain model we define three categories of devices: parts, actuators and sensors.
We ensure the BeSpaceD {\tt Component} construct is used to define every device by reusing the {\tt Device} class defined for the industrial automation domain.
{\small
\begin{verbatim}
    trait FestoDevice extends Device[ID]
\end{verbatim}
}

\indent We define the identity type for all devices as a String and use a type alias called {\tt ID}.

{\small
\begin{verbatim}
    type ID = String
\end{verbatim}
}

\indent This type is used consistently throughout the model where identity is required.
Finally, we define the three sub devices for this domain:

{\small
\begin{verbatim}
    class FestoPart(override val id: ID)
          extends Component(id) with Part[ID] with FestoDevice
    class FestoActuator(override val id: String)
          extends Component(id) with Actuator[ID] with FestoDevice
    class FestoSensor(override val id: ID)
          extends Component(id) with Sensor[ID] with FestoDevice
\end{verbatim}
}

\subsection{Device Descriptions }
\label{ssec:app-desc}
To construct device descriptions realizing meta data in the domain model we re-use the BeSpaceD {\tt ComponentDescription} class and refine the constructor parameter to be a behavioral map:

{\small
\begin{verbatim}
    class FestoComponentDescription(entries: BeMap[Component[Any],
                                                   ComponentValue[Any]])
          extends ComponentDescription[Component[Any],ComponentValue[Any]]
                  (entries.terms)
\end{verbatim}
}

Each device category requires different meta-data.
First we introduce a few concepts in Section~\ref{sssec:app-desc-anat}.
Then we define the data common to all devices in Section~\ref{sssec:app-desc-all}
followed by the meta data specific to each device category in its own section.

\subsubsection{Anatomy of a Part Description}
\label{sssec:app-desc-anat}
To illustrate the structure of the device descriptions we will briefly delve deep into one simple example; the StackEjector part.
(the {\tt "==>"} symbol is a binary infix operator for the BeSpaceD {\tt IMPLIES} construct).
Here is how we define it at the top level:

{\small
\begin{verbatim}
    val exampleEntry =  Component(StackEjector)  ==>
                        FestoComponentDescription(isPart, isHorizontalPusher,
                        stackEjectorVariations, stackEjectorOccupy)
\end{verbatim}
}

And this is what the parameters mean and how we construct them:
The device category and type is meta data about the type of the device and are simply key-value pairs constructed using an IMPLIES construct.

{\small
\begin{verbatim}
    val isPart               = DeviceCategory ==> Part
    val isHorizontalPusher   = DeviceType     ==> HorizontalPusherPart
\end{verbatim}
}

Spatial variations represent a discrete set of spatial positions that a part can be ``in''.
First, we define the set of demonstrator specific positions, and then map this to a mutual exclusion structure of component values.
We use the BeSpaceD exclusive or {\tt XOR} logical construct to represent mutual exclusion.

{\small
\begin{verbatim}
    val StackEjectorPositions = XOR(StackEjectorRetractedPosition,
                                    StackEjectorExtendedPosition
                                    )
    val stackEjectorVariations = SpatialVariations ==>
                                     ComponentValue(XOR(
                                         StackEjectorPositions.terms map 
                                             {pos => ComponentValue.apply(pos)}
                                         )
                                     )
\end{verbatim}
}

\begin{figure*}
    \centering
    \includegraphics[width=0.5\textwidth] {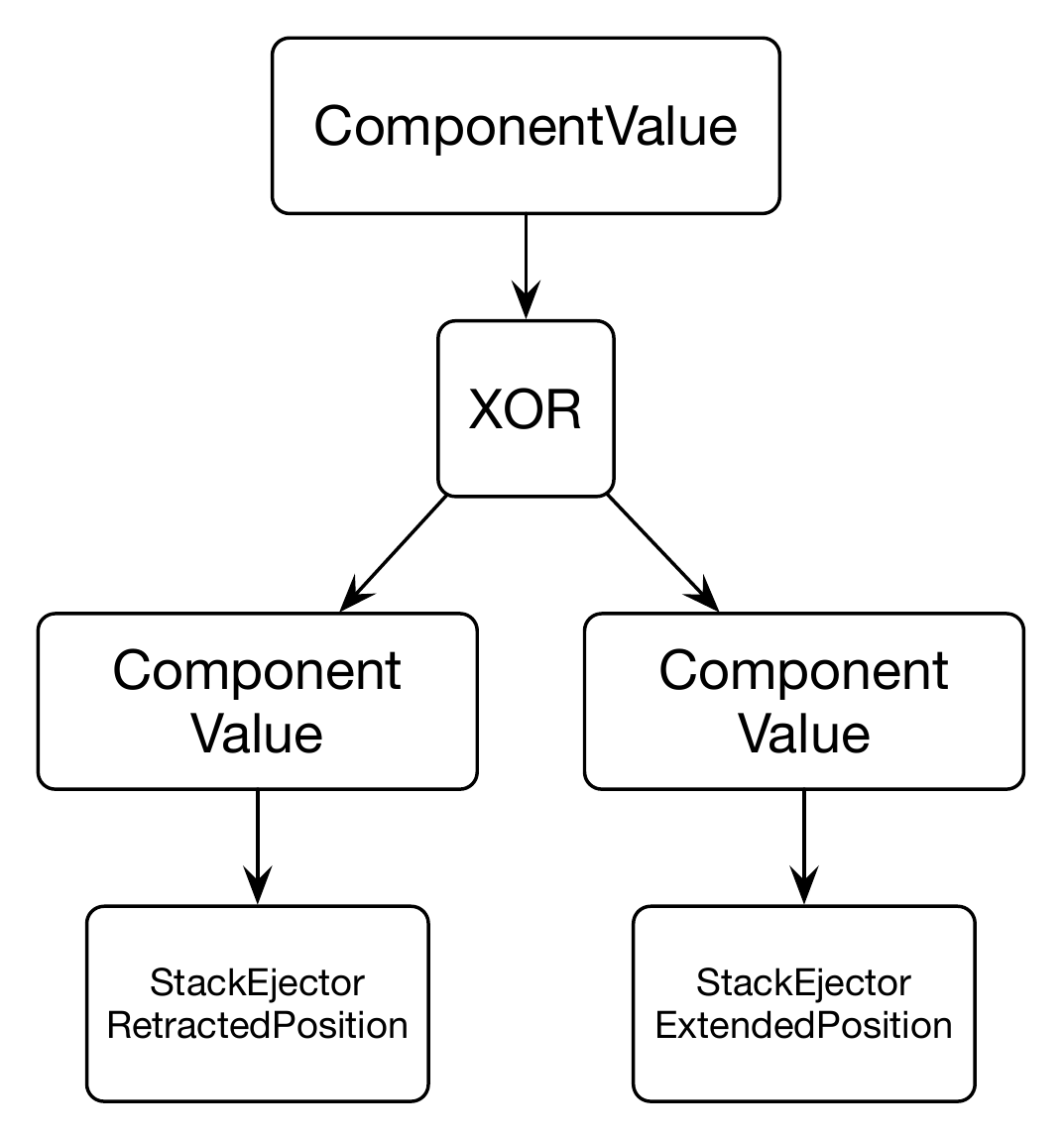}
    \caption{Spatial variations as a BeSpaceD object graph}
    \label{fig:variations}
\end{figure*}

\noindent
Figure~\ref{fig:variations} represents the above spatial variations as a graph of BeSpaceD objects.
The last parameter, {\tt stackEjectorOccupy}, defines the three dimensional space which the part occupies:
{\small
\begin{verbatim}
    val stackEjectorOccupy = SpatialLocation ==>
          ComponentValue(Occupy3DBox(
             X.StackEjectorLeft, Y.StackEjectorFront, Z.StackEjectorBottom,
             Width.StackEjector, Depth.StackEjector, Height.StackEjector
          ))
\end{verbatim}
}
\noindent The parameters in the {\tt Occupy3DBox} above are constant measurements.
How we define these measurements is discussed further in subsection~\ref{ssec:space}.

\subsubsection{Overview of all Device Descriptions}
\label{sssec:app-desc-all}
The meta data for multiple devices includes the keys: DeviceCategory, DeviceType, GPIO, PartAssociation, SignalMapping and SpatialLocation. Well known values are defined for the first three keys; for example:
{\small
\begin{verbatim}
    // Device Categories
    val Part           = ComponentValue("Part")
    val Actuator       = ComponentValue("Actuator")
    val Sensor         = ComponentValue("Sensor")
  
    // Device Types
    val SolenoidActuator = ComponentValue("Solenoid")
    val LightSensor = ComponentValue("Light Sensor")
  
    // GPIO Pins - Actuators
    val GPIO_1  = ComponentValue(1)
    val GPIO_26 = ComponentValue(26)
    
    // GPIO Pins - Sensors
    val GPIO_7  = ComponentValue(7)
    val GPIO_25 = ComponentValue(25)
\end{verbatim}
}

\noindent The GPIO Pins are either used for a sensor or an actuator, not both.

\subsubsection{Signal Mapping}
\label{sssec:sigmap}
Our factory demonstrator comes with a variety of actuators and sensors.
In station one, they all have binary signalling which means there are only two voltage levels generated or accepted.
These signals are named {\tt High} (for a high voltage level) and {\tt Low} (for a low voltage level).
It would be perfect if every actuator interpreted High as activation
and every light sensor generated a High signal for an obstructed state.
Unfortunately, this semantic mapping is inconsistent between devices, even in station one of our demonstrator factory.
Some solenoid based actuators activate on {\tt High} and some on {\tt Low}.
Some light sensors when obstructed generate {\tt High} and others {\tt Low}.
So we need a functional mapping between signal level and its meaning.

We encode the meaning of the signal to/from each actuator/sensor in a signal mapping data structure.
To model the meaning, we use the {\tt State} types defined in the industrial automation domain model (see Section~\ref{ssec:ia-var}).
The embedded object is of any type and allows us to create domain specific types for the demonstrator domain.
In our case we created the following signal types:

{\small
\begin{verbatim}
    class FestoSignal(val state: String)
    case object High extends FestoSignal("High")
    case object Low  extends FestoSignal("Low" )
    case object DC   extends FestoSignal("Don't Care" )
\end{verbatim}
}

\noindent And them embed in the device states:

{\small
\begin{verbatim}
    class FestoActuatorState(signal: FestoSignal)
          extends FestoDeviceState(signal) with ActuatorState[FestoSignal]
    class FestoSensorState(signal: FestoSignal)
          extends FestoDeviceState(signal) with SensorState[FestoSignal]
\end{verbatim}
}

\noindent The {\tt DC} signal is used to specify an abstracted device state where we don't care what the voltage signal actually is or will be.
In the application domain specific model, constants are defined for all the device states,
for example actuators that use a solenoid are defined as follows:

{\small
\begin{verbatim}
    abstract class FestoSolenoidActuatorState(signal: FestoSignal)
             extends FestoActuatorState(signal)
    class ActiveState(sig: FestoSignal)
          extends FestoSolenoidActuatorState(sig)
    class PassiveState(sig: FestoSignal)
          extends FestoSolenoidActuatorState(sig)

    val ActiveHigh  = ActiveState(High)
    val ActiveLow   = ActiveState(Low)
    val Active      = ActiveState(DC)
    val PassiveHigh = PassiveState(High)
    val PassiveLow  = PassiveState(Low)
    val Passive     = PassiveState(DC)
\end{verbatim}
}

\noindent Objects of these types are used in the mapping to describe the meaning of the signal.
The signal mapping function is defined as:
:
{\small
\begin{verbatim}
    type SignalTo[R]           = Map[FestoSignal, R]
    type SignalSensorMapping   = SignalTo[FestoSensorState]
    type SignalActuatorMapping = SignalTo[FestoActuatorState]
\end{verbatim}
}

\noindent and used for example in the following way:

{\small
\begin{verbatim}
    val HighSolenoidActuatorMapping: SignalActuatorMapping =
            Map(High -> ActiveHigh,  Low -> PassiveLow)  

    val SignalMapping = Component("Signal Mapping")
    val isHighSolenoidActuator =
            SignalMapping ==> ComponentValue(HighSolenoidActuatorMapping)
\end{verbatim}
}

\subsubsection{Parts}
The meta data for parts includes the key SpatialVariations and an
example of a well known value is:

{\small
\begin{verbatim}
    val ConveyerPart         = ComponentValue("Conveyer")
    val isConveyer           = DeviceType ==> ConveyerPart
\end{verbatim}
}

\noindent A spatial variation defines a discrete set of positions for a moving part semantically as discussed in Section~\ref{sssec:app-desc-anat}.

\subsubsection{Actuators}
Actuators are defined in terms of the common keys, however, they do
have some specific well known values such as:

{\small
\begin{verbatim}
    val SolonoidActuator = ComponentValue("Solonoid")  
    val isSolonoid = DeviceType ==> SolonoidActuator
\end{verbatim}
}

\noindent An example description for the vacuum gripper actuator is:

{\small
\begin{verbatim}
    val forLoader = PartAssociation   ==> ComponentValue(Loader)
    Component(VacuumGrip) ==> FestoComponentDescription(
                              isActuator, isSolonoid, isGpio5, 
                              isHighSolonoidActuator, forLoader) 
\end{verbatim}
}

\noindent Note, the spatial occupation data has been left out.
A topic for future work would be how to model spatial occupation when it is not fixed relative to the machine but will be moving relative to how the part it is embedded in moves.
The {\tt isHighSolonoidActuator} is a signal mapping function (see Section~\ref{sssec:sigmap}) that declares that a high voltage activates the solenoid and a low voltage inactivates it.

\subsubsection{Sensors}
Sensors have some specific well known values such as:

{\small
\begin{verbatim}
    val LightSensor = ComponentValue("Light Sensor")
    val isLightSensor = DeviceType ==> LightSensor
\end{verbatim}
}

\noindent An example of a sensor description follows. It uses the same meta data
keys as the actuators:

{\small
\begin{verbatim}
    Component(StackEjectorExtended)  ==> FestoComponentDescription(
                                      isSensor, isLightSensor, isGpio0,
                                      isHighLightSensor, forStackEjector,
                                      stackEjectorExtendedOccupy)
\end{verbatim}
}

\noindent The {\tt stackEjectorExtendedOccupy} parameter is spatial data and is explained in the following Section~\ref{ssec:space}.

\subsubsection{Spatial Meta Data}
\label{ssec:space}
We use the {\tt Occupy3DBox} core construct to define the volume of space occupied by a device.
The desired result of an {\tt Occupy3DBox} is quite simple, however, the steps to construct one is not so simple.
we did this in an elegant way that minimises the number of measurements that need to be taken by reusing the measurements of key reference points and sizes.
Below is a partial extract of the measurement code and serves as an
example for the two light sensors in the proximity of the cap ejector
part:

{\footnotesize
\begin{verbatim}
    object X {
      val Station1EdgeLeft         = 0
      val StackEjectorRight        = Station1EdgeLeft + 85
      val ExtendRetractSensorRight = StackEjectorRight
      val ExtendRetractSensorLeft  = ExtendRetractSensorRight - Width.ExtendRetractSensor
    }
    object Y {
      val Station1EdgeFront        = 0
      val StackEjectorFront        = Station1EdgeFront + 76
      
      val ExtendSensorFront        = StackEjectorFront + 122
      val ExtendSensorBack         = ExtendSensorFront + Depth.ExtendRetractSensor
      
      val RetractSensorFront       = StackEjectorFront + 236
      val RetractSensorBack        = RetractSensorFront + Depth.ExtendRetractSensor
    }
    object Z {
      val Base                        = 0
      val StackEjectorBottom          = Base
      val ExtendRetractSensorBottom   = Base + 4
      val ExtendRetractSensorTop      = Base + 20
    }
    object Width {
      val ExtendRetractSensor = 32
    }
    object Depth {
      val ExtendRetractSensor = 10
    }
    object Height {
      val ExtendRetractSensor = Z.ExtendRetractSensorTop -
                                Z.ExtendRetractSensorBottom
    }
\end{verbatim}
}

 Note, that the width, depth and height metrics are reused for both sensors as are the x and z coordinates.
The y coordinate is different for each sensor and so is separated.
In addition,  reference metrics are used that are reused to build up the actual metric symbolically.
There are several layers of referencing including
the base of the station one,
the edge of a part and
the edge of sensor or actuator.
These layers are referenced by downstream layers to formulate the metrics.
Finally some meta data can be created for defining the spatial
occupation for the sensor's descriptions:

{\small
\begin{verbatim}
    val stackEjectorExtendedOccupy  = occupy(
                                         X.ExtendRetractSensorLeft,
                                         Y.ExtendSensorFront, 
                                         Z.ExtendRetractSensorBottom,
                                         Width.ExtendRetractSensor,
                                         Depth.ExtendRetractSensor,
                                         Height.ExtendRetractSensor)
                                         
    Component(StackEjectorExtended)  ==> FestoComponentDescription(
                                            isSensor, isLightSensor, isGpio0, 
                                            isHighLightSensor, forStackEjector,
                                            stackEjectorExtendedOccupy)
    
    val stackEjectorRetractedOccupy = occupy(
                                         X.ExtendRetractSensorLeft,
                                         Y.RetractSensorFront,
                                         Z.ExtendRetractSensorBottom,
                                         Width.ExtendRetractSensor,
                                         Depth.ExtendRetractSensor,
                                         Height.ExtendRetractSensor)
                                         
    Component(StackEjectorRetracted) ==> FestoComponentDescription(
                                            isSensor, isLightSensor, isGpio3,
                                            isHighLightSensor, forStackEjector,
                                            stackEjectorRetractedOccupy)

\end{verbatim}
}

\noindent Notice that the {\tt OccupyBox} objects are consistently built from the left, front and bottom coordinates
as well as the width, depth and height sizes for each sensor.

In this section we have seen step by step how spatial data for the volume occupied by a device
is built from a few measurements with layered symbolic referencing.

\subsection{Topologies}
So far we have only addressed static meta data oriented around components.
In this section we show how we modelled information in a topological structure.
We have developed several different but complimentary topologies
that illustrate the possibilities that BeSpaceD offers in modelling topological information
and these are presented in the following.

\subsubsection{Sensor Process Sequence}
\begin{figure*}
    \centering
    \includegraphics[width=1\textwidth] {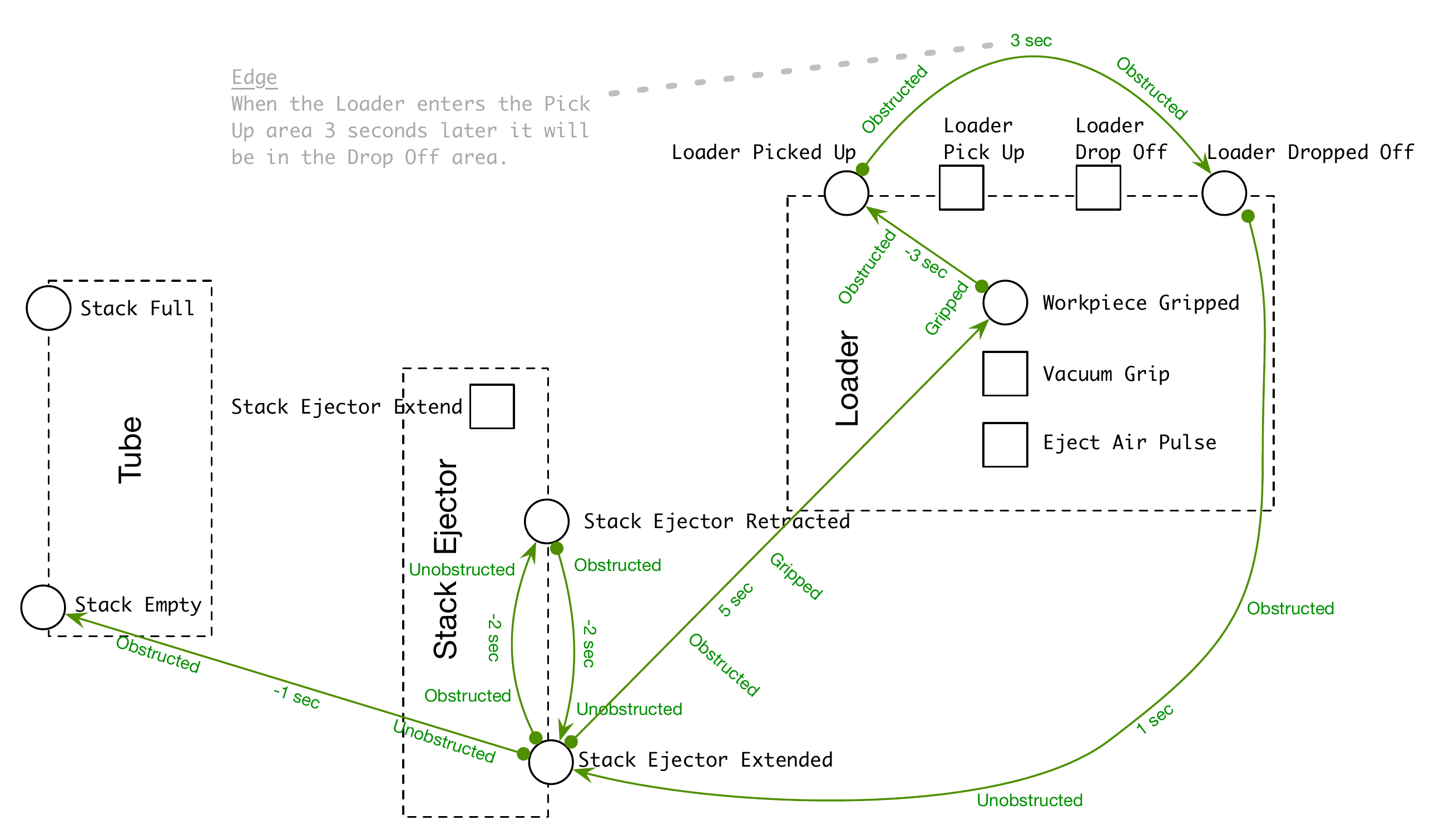}
    \caption{Process flow topology}
    \label{fig:top-procseq}
\end{figure*}

The first topology involves only sensors and captures the expected order of state change in a set of sensors as material flows through the machine during normal processing scenarios. The process flow topology is illustrated in Figure~\ref{fig:top-procseq}.

Each edge in the topology represents a temporal proximity relationship: A state change in one sensor (source node) is expected to be soon followed by a state change in another sensor (destination node). The way we model this is by representing sensors as nodes and the expectation as edges.
Each edge is annotated with three pieces of information:

{\small
\begin{itemize}
\item Cause: the state change of the first sensor. This is done by annotating the leading edge with the value the sensor state changes TO.
\item Duration: the exact time between the two state changes.
\item Effect: the state change of the second sensor. This is done by annotating the trailing edge with the value the sensor state changes TO.
\end{itemize}
}

\noindent The BeSpaceD encoding of the annotations use a {\tt
  TemporalCorrelation} as follows:

{\small
\begin{verbatim}
    class  FestoStateCorrelation(
              cause: FestoDeviceState,
              duration: TimeDuration[FestoDeviceState],
              effect: FestoDeviceState)
    extends TemporalCorrelation[FestoDeviceState, Integer]
            (cause, duration, effect)
\end{verbatim}
}

\noindent Below is an example of one such process sequence edge in BeSpaceD serialised to JSON
(this edge is the same one commented on in
Figure~\ref{fig:top-procseq}):

{\footnotesize
\begin{verbatim}
{
    "type" : "EdgeAnnotated",
    "source" : {"type":"Component","id":"Loader Picked Up"},
    "target" : {"type":"Component","id":"Loader Dropped Off"},
    "annotation" : {
        "type": "FestoStateCorrelation",
        "cause": {"type": "Obstructed"},
        "duration": {
            "type":"TimeDuration",
            "start":{"type": "SS Variable", "expression": {"type": "Obstructed"}},
            "scalar":{"type": "SS Addition", "expression": [
                     {"type": "SS Variable", "expression": {"type": "Obstructed"}},
                     {"type": "SS Constant", "expression": 3}
                     ]}
             },
         "effect": {"type": "Obstructed"}}
    }
\end{verbatim}
}

\noindent Notice that this topology does not allow for any variability in the timing so it would be best suited for modelling real world data inferred from live events.

\subsubsection{Actuator-Sensor Causality}
\label{sssec:cause}
\begin{figure*}
    \centering
    \includegraphics[width=1\textwidth] {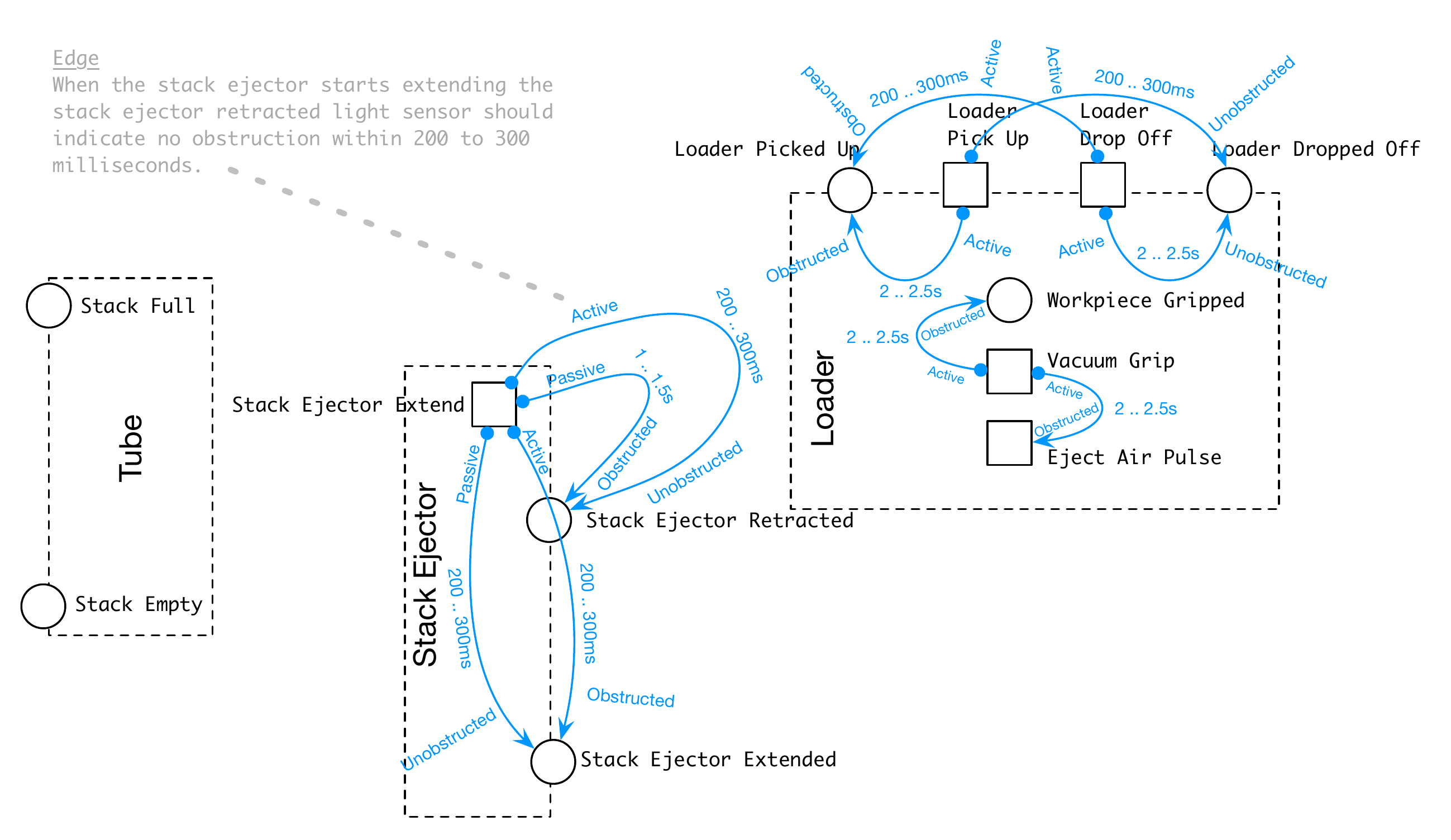}
    \caption{Causality topology}
    \label{fig:top-cause}
\end{figure*}

The causality topology is illustrated in Figure~\ref{fig:top-cause}.

Each edge in the causality topology represents a temporal proximity relationship between an actuator to a sensor: A state change in an actuator (source node) leads to a state change in a sensor (destination node). We assert that this must happen within some time range. The way we model this is by representing devices as nodes and the assertion as edges.
\indent Each edge is annotated with three pieces of information:

{\small
\begin{itemize}
\item Cause: the state change of the actuator. This is done by annotating the leading edge with the value the actuator state changes TO.
\item Duration Range: One minimum and one maximum time duration.
\item Effect: the state change of the sensor. This is done by annotating the trailing edge with the value the sensor state changes TO.
\end{itemize}
}

\noindent The BeSpaceD encoding of these annotations use a {\tt
  TemporalConstraint} as follows:

{\small
\begin{verbatim}
    class  FestoStateConstraint(
              cause: FestoDeviceState,
              durationRange: TimeDurationRange[FestoDeviceState],
              effect: FestoDeviceState, inverse: Boolean = false
              )
          extends TemporalConstraint[FestoDeviceState, Integer]
              (cause, durationRange, effect, inverse)
\end{verbatim}
}

\noindent Below is an example of one such causality edge in BeSpaceD
serialized to JSON (this edge is the same one commented on in
Figure~\ref{fig:top-cause}):

{\footnotesize
\begin{verbatim}
{
    "type" : "EdgeAnnotated",
    "source" : {"type":"Component","id":"Stack Ejector Extend"},
    "target" : {"type":"Component","id":"Stack Ejector Retracted"},
    "annotation" : {
        "type": "FestoStateConstraint",
        "cause": {"type": "Active"},
        "durationRange": {
            "type":"TimeDurationRange",
            "minimum":{
                "type":"TimeDuration",
                "start":{
                    "type": "SS Variable",
                    "expression": {"type": "Active"}},
                "scalar":{
                    "type": "SS Addition",
                    "expression": [
                        {"type": "SS Variable", "expression": {"type": "Active"}},
                        {"type": "SS Constant", "expression": 200}
                        ]}
                },
            "maximum":{
                "type":"TimeDuration",
                "start":{"type": "SS Variable", "expression": {"type": "Active"}},
                "scalar":{
                    "type": "SS Addition",
                    "expression": [
                        {"type": "SS Variable", "expression": {"type": "Active"}},
                        {"type": "SS Constant", "expression": 300}
                        ]}
                }
            },
    "effect": {"type": "Unobstructed"}}
}
\end{verbatim}
}

\subsubsection{Actuator-Actuator Avoidance (Safety Rules)}

\begin{figure*}
    \centering
    \includegraphics[width=1\textwidth] {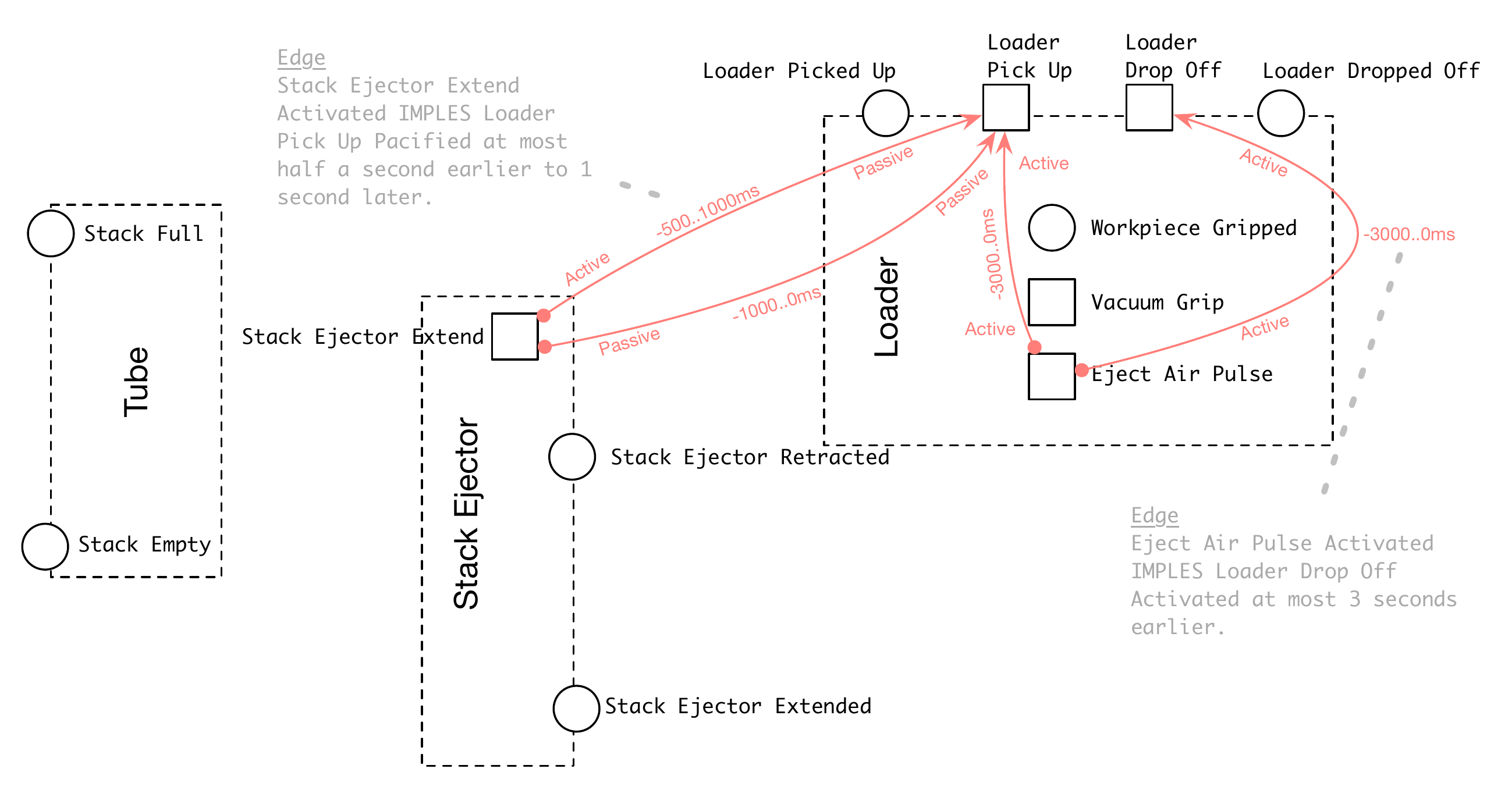}
    \caption{Avoidance topology}
    \label{fig:top-avoid}
\end{figure*}

The avoidance topology is illustrated in Figure~\ref{fig:top-avoid}.
Each edge in the avoidance topology represents a temporal proximity relationship between two actuators: A state change in a first actuator (source node) leads to a state change in a second actuator (destination node). We assert that this must happen within some time range. The way we model this is by representing devices as nodes and the assertion as edges.

{\small
\begin{itemize}
\item Cause: the state change of the first actuator. This is done by annotating with the leading edge with the value the actuator state changes TO.
\item Duration Range: One minimum and one maximum time duration.
\item Effect: the state change of the second actuator. This is done by annotating with the leading edge with the value the actuator state changes TO.
\end{itemize}
}

\noindent The BeSpaceD encoding of these annotations use a {\tt TemporalConstraint} as in Section~\ref{sssec:cause}.
Below is an example of one such avoidance edge in BeSpaceD serialized to JSON (this edge is the top edge commented on in Figure~\ref{fig:top-avoid}):

{\footnotesize
\begin{verbatim}
{
    "type" : "EdgeAnnotated",
    "source" : {"type":"Component","id":"Stack Ejector Extend"},
    "target" : {"type":"Component","id":"Loader Pickup"},
    "annotation" : {
        "type": "FestoStateConstraint",
        "cause": {"type": "Active"},
        "durationRange": {
            "type":"TimeDurationRange",
            "minimum":{
                "type":"TimeDuration",
                "start":{"type": "SS Variable", "expression": {"type": "Active"}},
                "scalar":{"type": "SS Addition", "expression": [
                    {"type": "SS Variable", "expression": {"type": "Active"}},
                    {"type": "SS Constant", "expression": -500}
                    ]}
                },
            "maximum":{
                "type":"TimeDuration",
                "start":{
                    "type": "SS Variable",
                    "expression": {"type": "Active"}},
                "scalar":{
                    "type": "SS Addition",
                    "expression": [
                        {"type": "SS Variable", "expression": {"type": "Active"}},
                        {"type": "SS Constant", "expression": 1000}
                        ]
                    }
                }
            },
    "effect": {"type": "Passive"}}
}
\end{verbatim}
}

\subsection{Events}
\begin{figure*}
    \centering
    \includegraphics[width=1\textwidth] {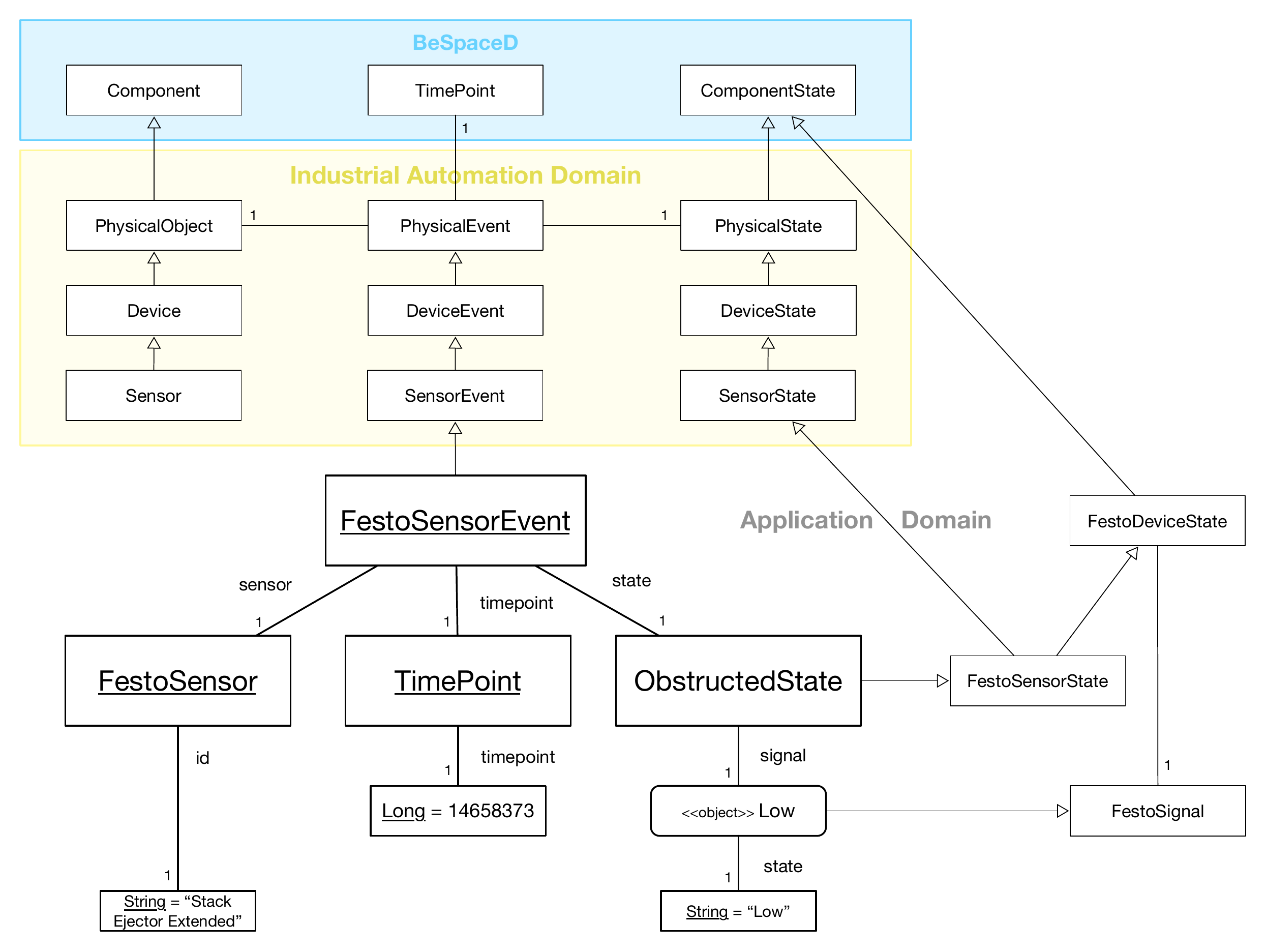}
    \caption{Example sensor event}
    \label{fig:event}
\end{figure*}

The model for live events from the food processing factory is built in layers of domain specific types that ultimately derive from BeSpaceD constructs.
Figure~\ref{fig:event} illustrates the BeSpaceD data structure for live events.
The figure also shows the traits from the industrial automation domain to further illustrate the layers.
The entity {\tt Low} on the diagram is an immutable singleton object
(in Scala) meaning the name refers to an object instance of a class
with the same name as the object.

{\small
\begin{verbatim}
    case object Low  extends FestoSignal("Low" )
\end{verbatim}
}

Types for sensors, actuators and parts are built from {\tt Component} objects.
Types for the state of these devices are built from {\tt ComponentState} objects.
And finally types for events are built from the core type in the industrial automation domain called {\tt PhysicalEvent}
(see Section~\ref{ssec:ia-event}).

Physical events are a composition of the previous two types plus a time point.
The three key pieces of information are
the object this event is scoped to,
the time this event occurred and
the state this object changed to.
Below is an example of how to create a live event for the demonstrator
plant.
Apart from the time this constrction code matched the the illustration in Figure~\ref{fig:event}:

{\small
\begin{verbatim}
      val now = new GregorianCalendar()
      val timePoint = TimePoint(now.getTime.getTime)
      val event = FestoSensorEvent(
                         CapDispenser.StackEjectorExtended,
                         timePoint,
                         ObstructedLow
                         )
\end{verbatim}
}

\noindent The state {\tt ObstructedLow} is predefined in our application domain
along with general light sensor states as follows:

{\small
\begin{verbatim}
      abstract class FestoLightSensorState(sig: FestoSignal)
               extends FestoSensorState(sig)
      class ObstructedState(sig: FestoSignal)
            extends FestoLightSensorState(sig)
      val ObstructedHigh = ObstructedState(High)
      val ObstructedLow = ObstructedState(Low)
\end{verbatim}
}

\subsection{Visualizing Topology and Events}
\begin{figure*}
    \centering
    \includegraphics[width=0.7\textwidth] {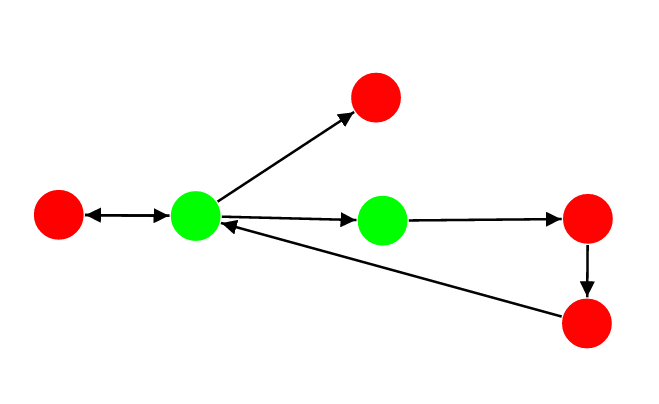}
    \caption{Example visualisation of topology and events}
    \label{fig:vis-basic}
\end{figure*}

We created a visualisation of live events from our food processing demonstrator in the context of a topology.
A snapshot of an example visualisation can be seen in Figure~\ref{fig:vis-basic}.

In this visualisation, a red node means the sensor is obstructed and a green node means it is not.
This corresponds to the the state changes inferred from the incoming live events.
The tool eStoreEd  was used to create this visualisation.

\section{Conclusion}
\label{sec:concl}
We have presented domain specific constructs for industrial automation
in our BeSpaceD modelling and reasoning framework. Furthermore, we
presented some specific insights into the formalization of our food
processing facility. This report continues a series of work, where we
exemplify and highlight characteristics of our BeSpaceD framework.

\subsection*{Acknowledgement}
We would like to thank Ian D. Peake, Lasith Fernando, James Harland, Huai Liu, Zoran Savic, and Yvette Wouters as well as our
students for supporting us with the formalization work of the factory demonstrator.

%------------------------------------------------------------------------------- Bibliography

 \bibliographystyle{eptcs}

\end{document}